# Electricity Price Forecasting: The Dawn of Machine Learning

by Arkadiusz Jędrzejewski[1], Jesus Lago[2], Grzegorz Marcjasz[1], Rafał Weron[1]

Electricity price forecasting (EPF) is a branch of forecasting on the interface of electrical engineering, statistics, computer science, and finance, which focuses on predicting prices in wholesale electricity markets for a whole spectrum of horizons. These range from a few minutes (real-time/intraday auctions and continuous trading), through days (day-ahead auctions), to weeks, months or even years (exchange and over-the-counter traded futures and forward contracts). Day-ahead (DA) markets are the workhorse of power trading, particularly in Europe, and a commonly used proxy for "the electricity price". The vast majority – up to 90% – of the EPF literature has focused on predicting DA prices. The latter are typically determined around noon during 24 uniform-price auctions, one for each hour of the next day. This has direct implications for the way EPF models are built.

Over the last 25 years, various methods and computational tools have been applied to intraday and day-ahead EPF. Until the early 2010s, the field was dominated by relatively small linear regression models and (artificial) neural networks, typically with no more than two dozen inputs. As time passed, more data and more computational power became available. The models grew larger to the extent where expert knowledge was no longer enough to manage the complex structures. This, in turn, led to the introduction of machine learning (ML) techniques in this rapidly developing and fascinating area. Here, we provide an overview of the main trends and EPF models as of 2022. Note that the article uses EPF as the abbreviation for both electricity price forecasting and electricity price forecast. The plural form, i.e., forecasts, is abbreviated EPFs.

# 25 Years of Evolution

The beginnings of EPF can be traced back to the 1990s. The first attempt to predict electricity price dynamics used linear regression techniques. Recall that such a model assumes a linear relationship between the predicted variable (e.g., the electricity price today at 6 p.m.) and the inputs (e.g., past electricity prices, load forecast for today). The inputs are also called features, explanatory variables, regressors, or predictors. More formally, the predicted variable is represented as a weighted sum of the inputs.

---

[1] Department of Operations Research and Business Intelligence, Wrocław University of Science and Technology, Poland
[2] Amazon, The Netherlands. Jesus Lago contributed to this work as an outside activity and not as part of his role at Amazon.

Early regression models were built on expert knowledge. The inputs included past prices (typically yesterday's, two and seven days ago), exogenous variables, and a seasonal component. Seasonality was either captured by a sinusoidal function or – more commonly – a set of so-called dummy variables taking the value of 1 on a given day of the week and 0 otherwise. A sample regression EPF model is illustrated in **Figure 1(a)**. White squares represent individual explanatory variables (also called regressors, inputs, or features), whereas the purple circle represents an output variable: $p_{d,h}$ and $X_{d,h}$ respectively stand for the electricity price and an exogenous variable (e.g., the day-ahead load forecast) for day $d$ and hour $h$, while $D_{Mon}, \ldots, D_{Sun}$ are the so-called dummy variables, with $D_{Mon} = 1$ for Monday and 0 otherwise, etc. Arrows indicate the flow of information – there is a coefficient (or weight) assigned to each arrow. The output is just a weighted sum of all the inputs.

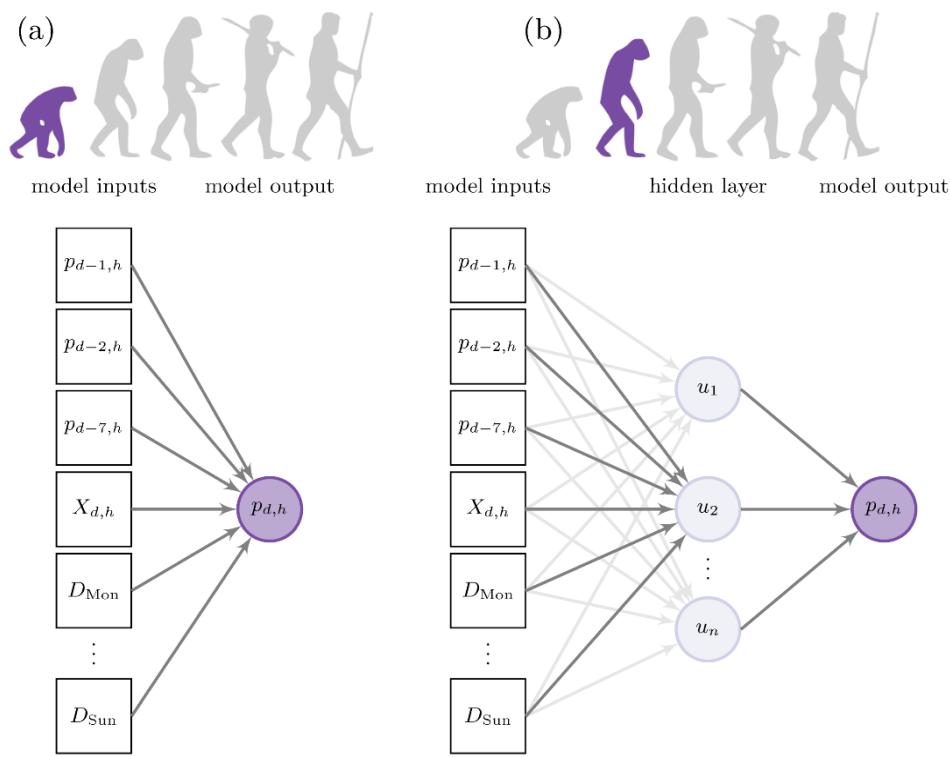

*Figure 1.* Early EPF models: *(a)* linear regression and *(b)* a single-output shallow neural network. White squares represent inputs, purple circles output variables and light blue circles hidden neurons (or nodes). Arrows indicate the flow of information (darker/lighter for better visibility).[3]

The most relevant and often the only used exogenous variable, even in more recent models, is the day-ahead load forecast. Interestingly, the actual load is not a better predictor. The reason is that the bids in the DA market are placed based on day-ahead load forecasts instead of the actual loads observed a day later. Since the electricity price-load relationship is nonlinear, see **Figure 2**, forecasters also played around with (artificial) neural networks.

---

[3] The human evolution graphic in Figures 1, 3, and 5 is a modification of the "Human Evolution Scheme" by José-Manuel Benitos, licensed under the terms of the GNU Free Documentation License.

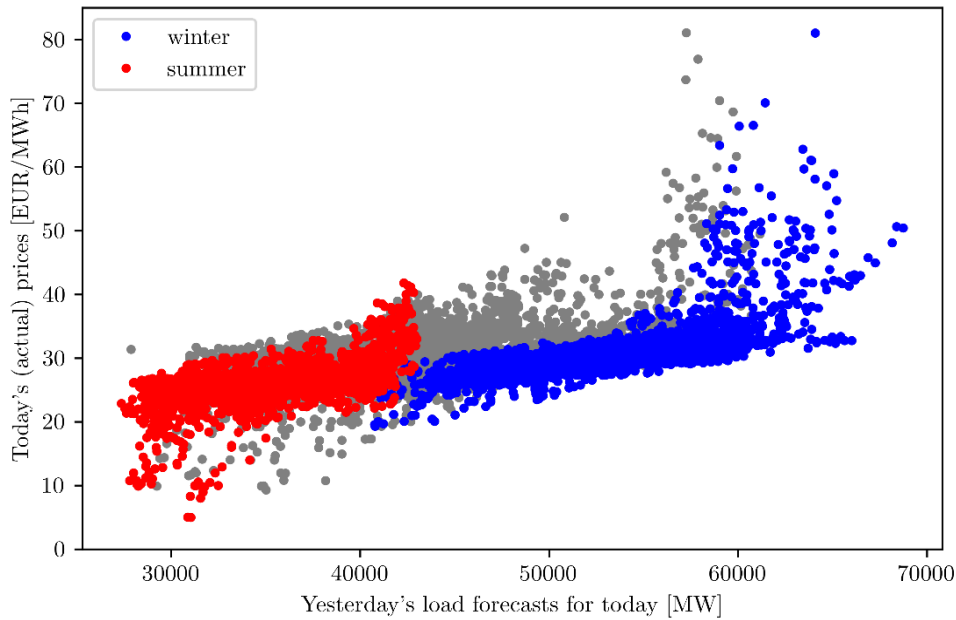

***Figure 2.*** *The relationship between electricity prices and the forecasted load is nonlinear and exhibits distinct seasonal variations, as illustrated by the 2017 data from the Scandinavian Nord Pool market.*

A neural network (NN) consists of layers of nodes. Each node is a neuron. Layers are connected by links between neurons, so a neuron can transmit a signal to another neuron in the subsequent layer, the structure of which resembles the synapses in a human brain. Thus, the output of a neuron is a weighted sum of all the inputs – as in linear regression – transformed by the so-called activation function. The simplest NN with input nodes connected directly to an output node with a linear activation function is equivalent to the linear regression model depicted in **Figure 1(a)**. By using nonlinear activation functions and inserting additional (so-called hidden) layers of nodes, the relationship between the inputs, i.e., the predictors, and the output, i.e., the predicted variable, becomes more complex. The NN models of the 1990s and the 2000s were typically shallow structures with only one hidden layer, one output neuron (e.g., the electricity price today at 6 p.m.) and a feed-forward architecture, see **Figure 1(b)**. The latter means that the information was propagated in only one direction – from inputs to the output. In contrast to linear regression where the output is just a weighted sum of all the inputs, in neural network models additional nonlinear transformations may be applied to this sum at each node.

## Shrinking Redundant Features

Uncovering the nonlinear relationship between the variables is a challenge for the linear regression based approach. Another challenge is the handling of a large number of inputs. Even a few dozens of explanatory variables can lead to unreliable estimates and predictions when using the classical regression model and ordinary least squares for minimizing the sum of squared errors between the forecasted and observed values. A working remedy provides so-called regularization algorithms. They shrink (hence the alternative name – shrinkage algorithms) coefficients, or weights of the less important inputs towards zero by adding a penalty for large coefficients to the sum of squared errors.

One of the most widely used regularization algorithms is the least absolute shrinkage and selection operator (LASSO), popularized by Robert Tibshirani in the mid-1990s. This machine learning (also called statistical learning) technique shrinks some coefficients to zero and thus performs semi-automatic, data-driven variable selection. EPF models that initially have hundreds of inputs can be reduced to structures with only a dozen or two relevant regressors. The selection is "semi-automatic" because LASSO requires setting the tuning hyperparameter that regulates which features are eliminated. This hyperparameter can be set using cross-validation, a resampling method that uses different portions of the data to test and train a model.

LASSO-estimated EPF models appeared in the mid-2010s and quickly revolutionized the field. Despite their relative simplicity and the inability to handle nonlinear dependencies, it is not easy to outperform them. Even using much more complex models. In one of the first EPF applications, LASSO was used to sparsify 24 regression models, one for each hour of the day and each with 100+ inputs, as shown in **Figure 3**.

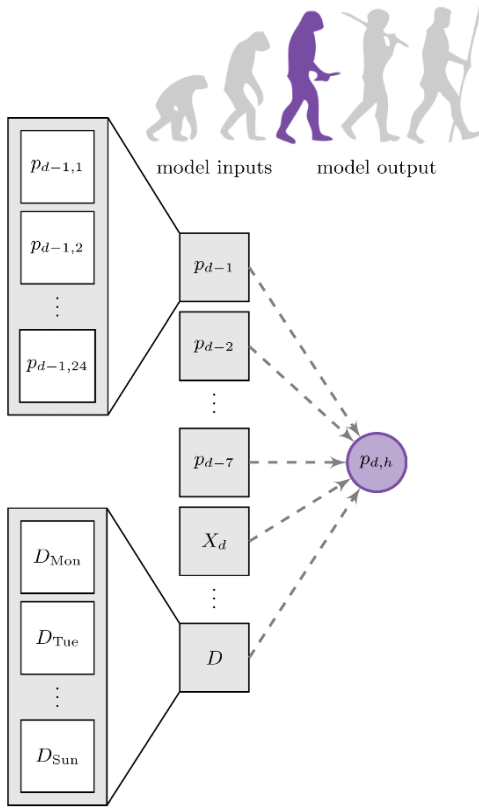

*Figure 3. Illustration of a LASSO-estimated regression, the next step in the evolution of EPF models. Grey squares denote vectors of variables and dashed arrows reflect the fact that LASSO may eliminate some of the links. Like in Figure 1, white squares represent inputs and the purple circle the output variable.*

A more recent example is the LASSO-Estimated AutoRegressive (LEAR) model, an open-source method that was proposed in 2021 as one of two challenging benchmarks for contemporary electricity price forecasting. The LEAR model is a parameter-rich autoregressive structure with about 250 explanatory variables. That is, to predict the price for one hour the model utilizes hourly prices from yesterday, two, three, and seven days ago, hourly day-ahead forecasts of two exogenous variables (e.g., system load and wind power generation) for today, yesterday, and a week ago, and weekday dummies. The model structure is visualized in **Figure 4** together with a graphical summary of the importance of selected inputs for Nord Pool market data. Four years of data (2013-2016) was used for calibration and two (2017-2018) for testing; the coefficients (weights) were re-estimated every day. The grey squares in the middle represent 24-dimensional vectors of explanatory variables: hourly prices yesterday, two, three and seven days ago ($p_{d-1}, p_{d-2}, p_{d-3}, p_{d-7}$) and hourly day-ahead forecasts of two exogenous variables for today, yesterday and a week ago ($X_d^i, X_{d-1}^i, X_{d-7}^i$; $i = 1$ for system load and $i = 2$ for wind power generation). The bottom grey square represents weekday dummies ($D$). The purple circle is the predicted variable – here the price for hour 18, i.e., 6 p.m. A set of 24 such models yields forecasts for all 24 hours of day $d$. The "chess-boards" to the left illustrate the importance of selected explanatory variables. The greener the square, the more often selected the feature (vertical axis) for a given target hour (horizontal axis; hour 18 is emphasized) in the test period. Dark grey squares represent features that were (almost) never selected. Clearly, there is a correlation between yesterday's load forecasts for hour $h$ today and the price for this hour – see the green diagonal in the middle "chess-board". In addition, yesterday's load forecasts for the morning peak (hours 8-9) and late evening (hours 22-24) also carry relevant information – see the green horizontal stripes in the same "chess-board".

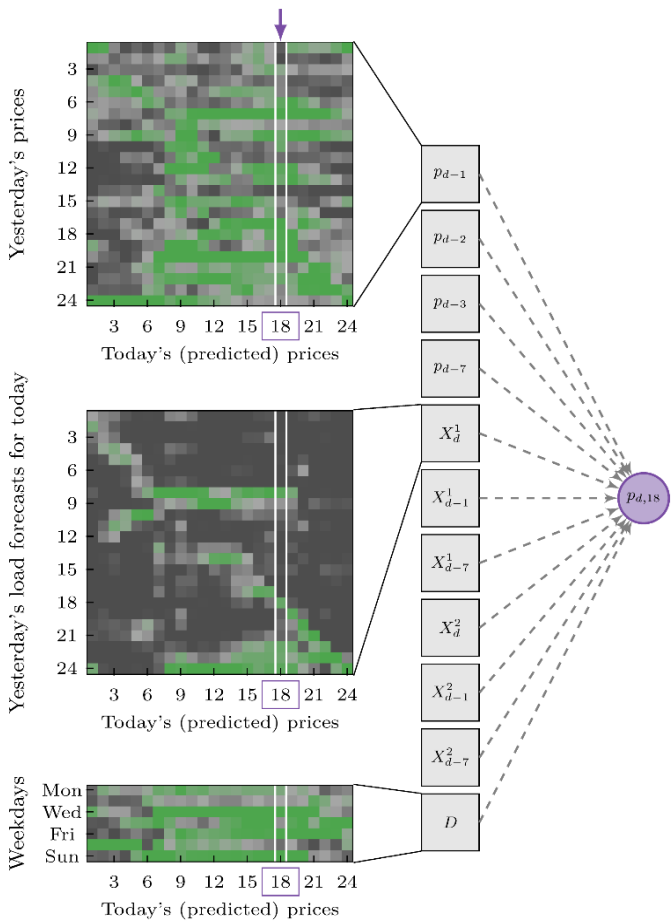

***Figure 4.*** *Visualization of variable selection in the LEAR (i.e., LASSO-Estimated AutoRegressive) model calibrated to data from the Nord Pool power market.*

## Deeper and Deeper

In the past, the challenge of using neural networks was computational complexity. Training small structures – like the one in **Figure 1(b)** – was feasible, but calibrating larger models was time/resource consuming and in some cases even impossible. For instance, during the optimization phase, the gradient would often become either zero or infinity, leading to the so-called vanishing gradient problem, effectively preventing the weights from changing their values.

Over the last decade, advances in computational resources (e.g., massive usage of graphics processing units, GPUs, and optimization algorithms) have made it possible to efficiently train complex structures, including neural networks with hundreds of inputs, multiple outputs, multiple hidden layers, and links to earlier layers, see **Figure 5**. As networks whose depth (i.e., the number of layers) was not just limited to a single hidden layer systemically showed better results and generalization capabilities, the field and models were named deep learning (DL) and deep neural networks (DNN) to stress the importance of depth in the achieved improvements.

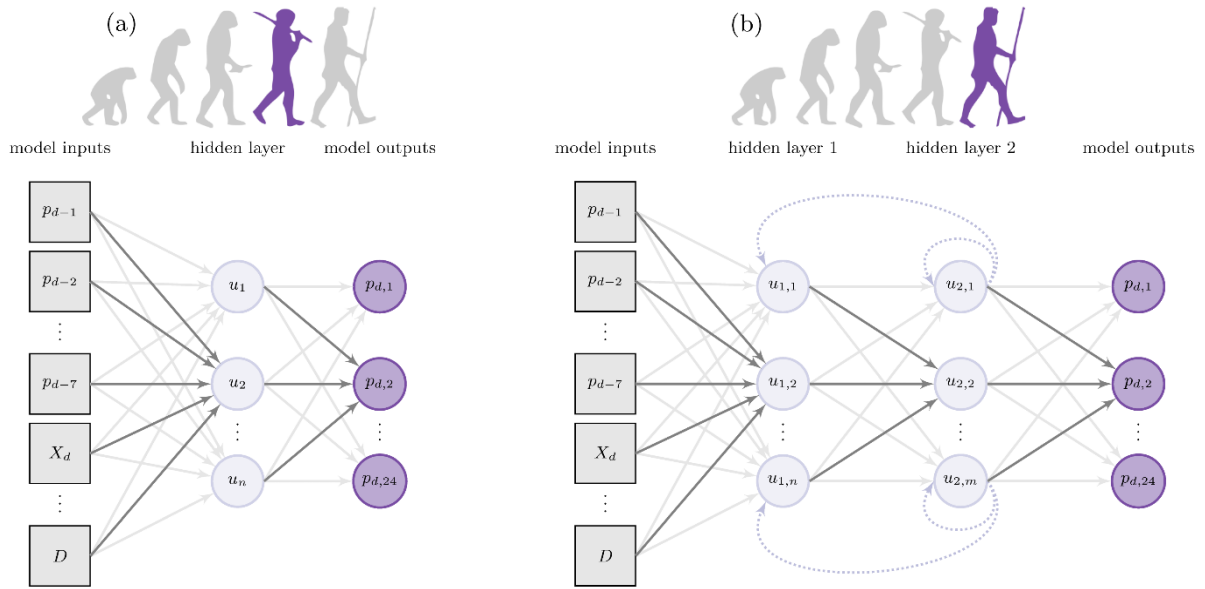

*Figure 5. Multi-output neural network architectures for EPF: **(a)** shallow and **(b)** deep with two hidden layers. The notation is similar to that in Figures 1 and 3. Dotted arrows represent feedback connections; if present, the structure is called a recurrent neural network (RNN).*

While this success of DL models initiated in computer science applications, e.g., image recognition, speech recognition, or machine translation, the benefits of DL also spread in the late 2010s to energy-related applications, e.g., electricity price or wind power forecasting. Since then, deep models have been heavily used to better exploit and model the nonlinear relations between energy-based quantities (prices, load, generation) and their drivers (e.g., human behaviour, calendar effects).

In the context of EPF, early works were based on feed-forward architectures with features (i.e., inputs) modeled as hyperparameters. Hyperparameters are model parameters that cannot be optimized during normal training. They are settings of the models (e.g., the number of neurons per layer, the type of activation function, the dropout rate, the learning rate, the regularization coefficient, the choice of optimization algorithm) that need to be optimized and tuned separately from the main NN parameters (i.e., the weights). Like hyperparameters, the features need to be selected/optimized separately from the main parameters. So, these early works proposed deep neural networks where the hyperparameters and inputs were optimized together.

The most prominent example of deep neural networks for EPF is probably the DNN model proposed in 2018. It is a DL model whose inputs and hyperparameters are optimized and tailored for each case study without the need for expert knowledge. It uses one of the simplest deep architectures proposed for EPF: feed-forward with two hidden layers, 24 outputs (i.e., it jointly predicts 24 hourly prices), and the same inputs as the LEAR model. DNN's structure is identical to the one in **Figure 5(b)**, except for the feedback connections (dotted arrows). In 2021, its open-source code was released and the DNN model itself was recommended as a challenging benchmark – even more than the LEAR model – for contemporary electricity price forecasting.

In terms of training and real-time usage, the DNN consists of two phases. In the first phase, the hyperparameters and input features are jointly optimized using historical data. To do so, the inputs are modeled as binary hyperparameters which can be either selected or discarded. This step is performed periodically but not very often, e.g., once per month. In the second phase, using the optimal set of inputs and hyperparameters, the DNN is recalibrated daily to account for the newest market data. Since each hyperparameter/feature selection yields a different local optimum, the DNN can greatly benefit from averaging forecasts. That is, training multiple DNNs and building a forecast as the arithmetic average of the individual predictions of the DNNs.

The first wave of DL models was followed in the last few years by hybrid architectures utilizing so-called long-short term memory and/or convolutional neural networks. Unlike feed-forward networks, the long-short term memory architecture has multiple feedback connections and can process not only single data points but also entire time series. Convolutional neural networks are regularized – to prevent overfitting – versions of multilayer feed-forward networks that use convolution (instead of matrix multiplication) in at least one of their layers.

However, despite the recent work on DL models, it is unclear whether all the extra complexity brings improvements in forecasting accuracy. The existing evidence indicates that, although simple DL models (like DNN) can on average improve upon LASSO-estimated regression models (like LEAR), the performance of more complex DL models is generally unknown. A common pattern that many works in DL share is the failure to compare the proposed models with state-of-the-art statistical methods and/or to employ long enough datasets to derive statistically sound conclusions.

## Model Performance

Let us now compare the predictive accuracy of the two challenging benchmarks – LEAR and DNN. Here, we report on a comprehensive study involving five major day-ahead electricity markets with test periods spanning two years each. Three out of five markets – Germany, France and Belgium – are operated by EPEX SPOT, the largest power exchange in Europe. In their case, the test periods range from 4 January 2016 to 31 December 2017 for Germany and 4 January 2015 to 31 December 2016 for France and Belgium. The remaining two include Nord Pool, one of the world's oldest power markets operating in Scandinavia and the Baltic states, and the Pennsylvania-New Jersey-Maryland (PJM) Interconnection, the largest competitive wholesale electricity market in the Americas. In their case, the test periods range from 27 December 2016 to 24 December 2018.

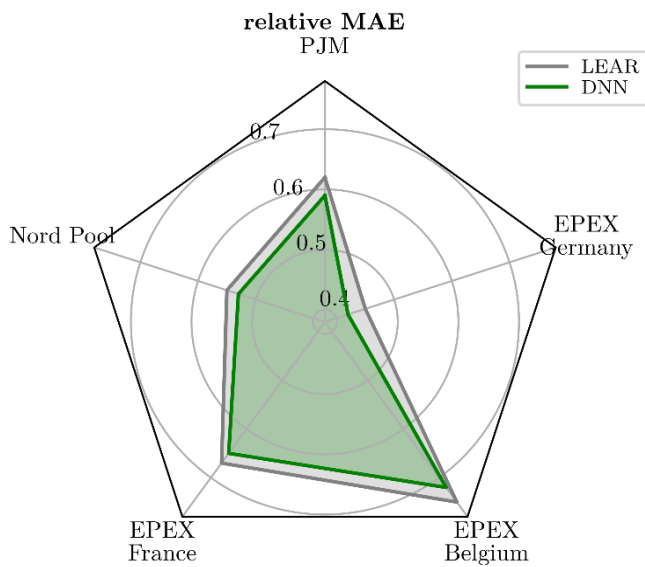

*Figure 6. Radar plot of the relative Mean Absolute Errors (relative MAE or rMAE) for the LEAR and DNN models and two-year test periods from five power markets.*

In **Figure 6** we plot the performance of a LEAR and a DNN model. The errors are computed with respect to a naive forecast which takes last week's price for the same hour. Both models provide significantly more accurate forecasts in terms of the mean absolute error (MAE) than the naive prediction; if plotted, the relative MAE (rMAE) of the naive method would be 1. For instance, for the PJM market, the MAE values of both models are lower by 40% (rMAE = ~0.6) than the MAE of the naive approach, while for the German EPEX market by as much as 55% (rMAE = ~0.45). Moreover, for all five electricity markets, the LEAR model is outperformed by the DNN. Clearly, there is a benefit of using the more complex method. However, it comes at a price – the significantly, up to 100 times higher computational cost. Is it worth the effort? Does it pay off? To answer these questions requires a much more thorough analysis, involving diverse error metrics, trading strategies, realistic operational costs, etc.

## Outlook for the Future

The methods presented so far yield point forecasts, like the expected price today at 6 p.m. This knowledge is important for decision-makers, but it does not convey all the relevant information. Probabilistic EPF is a possible remedy. The output of such a model will no longer be a single value, e.g., the expected price today at 6 p.m., but a distribution of possible values. For instance, looking at a probabilistic (also called distributional) forecast we will be able to conclude that the price today at 6 p.m. will be between 100 and 110 EUR with 90% probability.

There are multiple approaches to probabilistic forecasting. However, one particularly relevant in our case uses NNs to output the distributions directly. This can be achieved by making only a small change in the DNN model presented earlier. Instead of returning the 24 hourly prices, the network can return the parameter sets of 24 probability distribution functions. Assuming that the distribution of interest is Gaussian, the probabilistic NN will have 48 outputs with two parameters (i.e., the mean and standard deviation) for each hour of a day, as illustrated in **Figure 7**. Compared to the DNN in **Figure 5(b)**, the so-called distribution layer in **Figure 7** is placed between the last hidden and the output layer. Nodes of the distribution layer contain estimates (or predictions) of the parameters of a distribution (here: Gaussian). The output is a predictive distribution, $f_h$, one for each hour of the day.

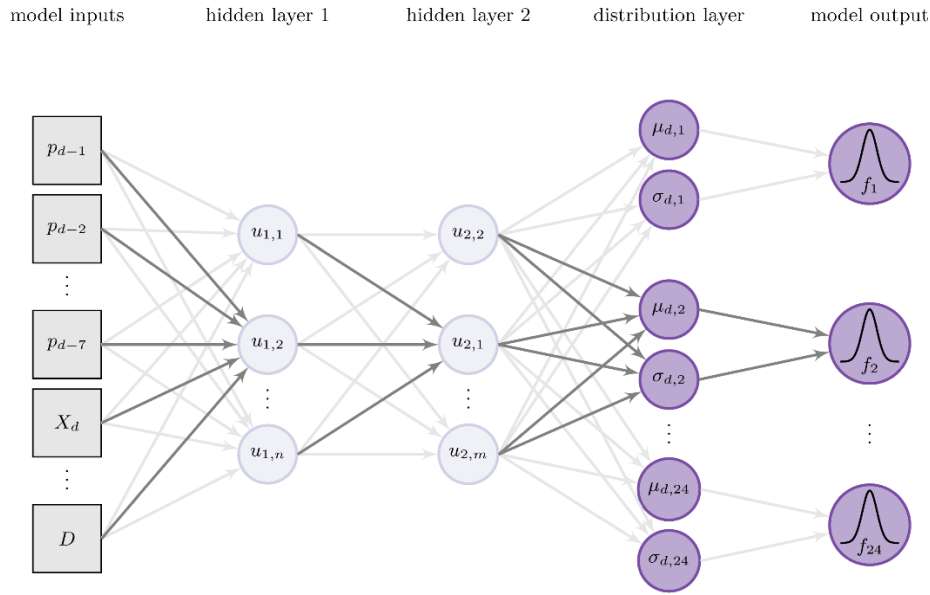

*Figure 7.* Visualization of a probabilistic neural network architecture for EPF. The output is a set of 24 predictive distributions, one for each hour of the day.

The downside is that the distribution itself is a hyperparameter, i.e., it has to be estimated or set ex-ante. A more flexible distribution can account for skewness and heavy tails, but this also comes at a price. In general, the more parameters there are to be estimated, the harder it is to calibrate the model to a given dataset. Nevertheless, probabilistic NNs may be objects of interest in the future.

# The Good, the Bad and the Ugly

Like the main characters in the epic spaghetti western film directed by Sergio Leone, the practices we can observe in the EPF literature come in all shapes and sizes. They range from highly undesirable (the bad, which lead to meaningless results that cannot be compared) and undesirable (the ugly, which compromise research reproducibility), to worth recommending (the good). We conclude this article by summarizing these practices – particularly common in the ML electricity price forecasting literature – and providing better alternatives.

## The Bad – Avoid at All Costs

New models are rarely compared against established state-of-the-art models, even less frequently against open-access benchmarks. This leads to unjustified claims regarding the "outstanding" performance of newly proposed models. From a practitioner perspective, such results are meaningless. The cost of implementing a new predictive model may be high, but its performance may be mediocre.

Another problem is the limited size and scope of the test dataset. Results are usually reported for just one market and often just comprise a selection of four weeks. This favors selectively picking results and leads to inflated forecasting accuracy.

The last issue is data contamination. The test dataset is not always selected as the last section of the full dataset nor is the test dataset completely independent of the validation/training datasets. Moreover, model hyperparameters are sometimes estimated in the test dataset.

## The Ugly – Handle with Care

The use of evaluation metrics is not standardized for the price forecasting community. New methods are often compared based on a single accuracy metric without considering its properties. Also, statistical testing is not always used to measure whether the differences in accuracy are statistically significant. This can lead to selectively picked results and biased conclusions.

Another issue is the lack of details when proposing new methods. In particular, the split between training, validation, and testing is often not reported nor are the optimal input features of the model or the optimal model hyperparameters. This hinders reproducibility. Moreover, the computational costs of new methods are often ignored. As short-term EPF often requires real-time forecasting, the computational cost is an important metric that should be considered.

Finally, models are not always recalibrated daily. This is particularly true for benchmarks. Often, a new proposed model would be frequently recalibrated, but the benchmark estimated only once and directly evaluated for the whole dataset. This leads to reporting worse accuracy and, when done for benchmarks, leads to artificial accuracy improvements of new models.

## The Good – Best Practices

Based on the issues listed above, new work in EPF should aim to follow a series of good practices to ensure reproducibility and meaningful comparisons. In particular, it should ensure that:
- The test dataset comprises at least a year of data and is based on multiple markets.
- The test dataset is selected as the last section of the available electricity price series and hyperparameters are estimated using a validation dataset that is different from the test dataset.
- Any new model is tested against well-known state-of-the-art, preferably open-sourced, models and well-known open-access datasets.
- To evaluate the models, several metrics are considered, and one of them is the relative MAE.
- Statistical testing is used to assess whether differences in predictive performance are significant.
- The split and dates of the dataset are explicitly stated, and all the inputs and parameters of the model are explicitly defined.
- The computational cost of new methods is evaluated and compared against the computational cost of existing methods.
- Forecasting models are recalibrated daily and not simply estimated once and evaluated in the full test dataset.

# Further Reading

[1] I. Goodfellow, Y. Bengio, A. Courville (2016) *Deep Learning*, MIT Press.
[2] R.J. Hyndman, G. Athanasopoulos (2021) *Forecasting: principles and practice* (3$^{rd}$ ed.), OTexts (https://otexts.com/fpp3/).
[3] G. James, D. Witten, T. Hastie, R. Tibshirani (2021) *An Introduction to Statistical Learning with Applications in R*, 2nd edition, Springer (https://www.statlearning.com).
[4] J. Lago, G. Marcjasz, B. De Schutter, R. Weron (2021) Forecasting day-ahead electricity prices: A review of state-of-the-art algorithms, best practices and an open-access benchmark, *Applied Energy* 293, 116983 (doi: 10.1016/j.apenergy.2021.116983). Python codes: https://epftoolbox.readthedocs.io.
[5] R. Weron (2014) Electricity price forecasting: A review of the state-of-the-art with a look into the future, *International Journal of Forecasting* 30(4), 1030-1081 (doi: 10.1016/j.ijforecast.2014.08.008).
[6] F. Ziel (2016) Forecasting electricity spot prices using LASSO: On capturing the autoregressive intraday structure, *IEEE Transactions on Power Systems* 31(6), 4977-4987.